\documentclass[showkeys,reprint,superscriptaddress,nofootinbib,amsmath,amssymb,aps,floatfix,]{revtex4-1}
\usepackage[latin1]{inputenc}
\usepackage{float}
\usepackage[english]{babel}
\usepackage{wasysym}
\usepackage{fancyhdr}
\usepackage{esdiff}
\usepackage{comment}
\usepackage{textcomp}
\usepackage{cancel}
\usepackage{tikz} 
\usepackage{subfigure}
\usepackage{graphicx}
\usepackage[font=small]{caption}
\usepackage{epsfig}
\usepackage{epstopdf}
\usepackage[T1]{fontenc}
\usepackage{lmodern}
\usepackage{ctable}
\usepackage[font=small,labelfont=bf,justification=raggedright]{caption}

\newcommand*{\avk}{\langle k \rangle}
\newcommand*{\bd}{\boldsymbol}
\newcommand{\Lp}{\mathcal{L}}

\begin{document}

\title{Diffusive behavior of multiplex networks}

\author{Giulia Cencetti}
\affiliation{Dipartimento di Ingegneria dell'Informazione, Universit\`{a} degli Studi di Firenze}
\affiliation{Dipartimento di Fisica e Astronomia, Universit\`{a} degli Studi di Firenze, INFN and CSDC}

\author{Federico Battiston}
\affiliation{Department of Network and Data Science, Central European University, Budapest 1051, Hungary}

\begin{abstract}
	Diffusion describes the motion of microscopic entities from regions of high concentration to regions of low concentration. In multiplex networks, flows can occur both within and across layers, and super-diffusion, a regime where the time scale of the multiplex to reach equilibrium is smaller than that of single networks in isolation, can emerge due to the interplay of these two mechanisms. In the limits of strong and weak inter-layer couplings multiplex diffusion has been linked to the spectrum of the supra-Laplacian associated to the system. 
	However, a general theory for the emergence of this behavior is still lacking. Here we shed light on how the structural and dynamical features of the multiplex affect the Laplacian spectral properties. For instance, we find that super-diffusion emerges the earliest in systems with poorly diffusive layers, and that its onset is independent from the presence of overlap, which only influences the maximum relative intensity of the phenomenon.
Moreover, a uniform allocation of resources to enhance diffusion within layers is preferable, as highly intra-layer heterogenous flows might hamper super-diffusion. Last, in multiplex networks formed by many layers, diffusion is best promoted by strengthening inter-layer flows across dissimilar layers.  
	 Our work can turn useful for the design of interconnected infrastructures in real-world transportation systems, clarifying the determinants able to drive the system towards the super-diffusive regime.

\end{abstract}	
 
\maketitle

\section*{Introduction}

Diffusion processes are widespread in nature and are known to be at the heart of many complex emerging collective behaviours, from biology to physics, such as contagions, animal migration, spreading of innovations, electric current in semiconductors, and Turing patterns~\cite{Gardiner09,Crank79,Murray02,Turing52}. A diffusion process is a macroscopic phenomenon resulting from the motion in space of microscopic entities, from regions of high concentration to regions of low concentration. In many cases of interest, it is natural to schematise the hosting spatial support as a discrete set of points, connected by means of an intricate skeleton of channels, which can be naturally represented as a complex network. In this framework, the microscopic agents move from node to node and the mobility process is governed by the fact that every discrete step is performed from each node towards the less crowded adjacent ones~\cite{Boccaletti06,BarratBarthelemyVespignani08,MasudaPorterLambiotte17,Strogatz01}. This results in a flowing mechanism which asymptotically ends when all the nodes are equally populated. Such homogeneous state, where the concentration of agents is uniformly distributed, represents a stable equilibrium for the system, and, differently from related mobility processes such as random walks~\cite{reactRW}, does not depend on the structural features of the network. Nevertheless, the topology of the interactions has an important effect on the transient dynamics of the system, ultimately setting the time scale needed to reach the eventual equilibrium.

In many real-world systems, from social~\cite{Szell10} to transportation networks~\cite{DeDomenico14navigability}, individual units can be connected through links which differ for meaning and relevance.
For instance, the underground, bus and railway networks coexist in many cities giving rise to multimodal transportation systems where each network is associated to a different spatial and temporal scale. These systems are well described by multiplex networks, where links of different type are embedded into separate layers of interactions~\cite{Boccaletti_etal14review,Kivela_etal14review,Battiston_etal17}. Diffusion processes were among the first dynamics introduced in the context of multiplex networks, where diffusion can occur both within and across layers~\cite{Gomez_etal13, Salehi_etal15,DeDomenico16, deArruda18fundamentals}. In this scenario, flows not only take place through nodes connected at a given layer, but also across two replicas of the same node belonging to two different layers. This is for instance the case of several main stations in major cities like London, where it takes time to move from the train platforms to those of the underground network, despite the two locations are both identified with the same name. It is possible to gain analytical insights on multiplex diffusion by studying the spectrum of the supra-Laplacian associated to the multi-layer system. Interestingly, Ref.~\cite{Gomez_etal13} showed that multiplex networks can have super-diffusive behavior, meaning that their time scale to relax to the steady state is smaller than that of any layer taken in isolation. Since then, the spectral properties of multiplex networks have been widely investigated~\cite{Sole_etal13, RadicchiArenas13, Sanchez_etal14, Cozzo16spectral, deArruda_etal18}, and the formalism has been extended to describe more complex phenomena, such as reaction-diffusion~\cite{Asllani_etal14,KouvarisHataDiaz15,BusielloCarlettiFanelli18} and synchronization processes~\cite{delGenio_etal16}.    
Nevertheless, while it was suggested that low correlation in the structure of the layers can enhance diffusion in a multiplex~\cite{Serrano_etal17}, a rigorous theory for the emergence of super-diffusion is currently lacking. 

In this work we unveil the main structural and dynamical determinants of multiplex super-diffusion. For instance, low overlap between the edges of multi-layer systems always maximizes the speed of the process when diffusion across layers is high. However, surprisingly, link correlations across layers do not affect positively the onset of super-diffusivity, which we find to be independent from edge overlap. 
In the past, the lack of such correlations has already been found responsible to increase the fragility of multi-layer systems~\cite{Cellai13}, 
to maximize the mixing of random walkers~\cite{Battiston16efficient},
 to promote multiculturality in the Axelrod model~\cite{Battiston17axelrod}, and hinder the beneficial effect of interconnectedness to cooperative games~\cite{BattistonPercLatora17}.

Moreover, we find that the faster the diffusive structure of the individual layers (higher density, broader degree distributions, etc), the lower the beneficial effect of multiplexity to the velocity of the process. Multiplex diffusion is promoted when the strength of diffusion within each layer is of the same order. For instance, super-diffusion might not be possible in the system if there exists at least one very slow layer, no matter the speed of diffusion across the other layers. Last, in multiplex networks composed by a large set of partially overlapping layers, diffusion is promoted by increasing the levels of interactions, and by preferentially enhancing inter-layer diffusion across networks with very different structure.

The paper is organized as follows. We first introduce the mathematical framework to describe diffusion in networks composed by many layers, summarize the main analytical results and define a novel indicator of multiplex super-diffusion. We then investigate the main structural and dynamical determinants of diffusion in a simple scenario focusing on multiplex networks with two layers only, both of them regular random graphs. This simple scenario allows us to isolate the effect of some important variables, namely the edge overlap between the layers, their size and average connectivity, and the inter- and intra-layer diffusion coefficients. We afterward extend our analysis to scale-free topologies, and the case of networks formed by a large number of layers. Finally we conclude by discussing possible further extensions of our work.

\section*{Model}
\label{sec_model}

Let us consider a multiplex network composed by $N$ nodes that can interact across $M$ different layers. The structure of the multiplex network can be described by associating an adjacency matrix $A^{[\alpha]} = \{ a_{ij}^{[\alpha]} \}$, $\alpha=1, \ldots, M$ to each layer, where $a_{ij}^{[\alpha]}=1$ if $i$ and $j$ are connected at layer $\alpha$, and $a_{ij}^{[\alpha]}=0$ otherwise. For the sake of simplicity, we consider the case where all connections are undirected.

The process of diffusion on this complex support can be studied by considering the time evolution of the state $x^{[\alpha]}_i$ of a generic flowing quantity on node $i$ at layer $\alpha$. The nodes are ordered according to the index $i+(\alpha-1)N$ with $i=1,...,N$, such that the complete state vector is $\bd x\in R^{N\times M}$. The diffusion equation reads:
\begin{equation}
 \begin{split}
 \dot x^{[\alpha]}_i =& D^{[\alpha]}\sum_{j=1}^{N}a^{[\alpha]}_{ij}(x^{[\alpha]}_j-x^{[\alpha]}_i)+\\ 
 &+ \sum_{\beta=1}^M D_x^{[\alpha,\beta]}(x^{[\beta]}_i-x^{[\alpha]}_i),
   \end{split}
 \label{diff}
\end{equation}
where $D^{[\alpha]}$ is the diffusion coefficient within each layer $\alpha$. Within nodes, diffusion can also occur across layers, and it is mediated by an inter-layer coefficient $D_x^{[\alpha,\beta]}$, with $D_x^{[\alpha,\alpha]}=0$ $\forall \alpha$~\cite{Gomez_etal13,Sole_etal13}. In matricial form Eq.~\ref{diff} takes the form: 
\begin{equation*}
\dot{\bd x} = - \Lp^{\mathcal M} \bd x
\end{equation*}
where $\Lp^{\mathcal M} \in R^{NM\times NM}$ denotes the supra-Laplacian defined in~\cite{Gomez_etal13,Sole_etal13} as:
\begin{equation*}
 \Lp^{\mathcal M} = \Lp^{\ell} + \Lp^x.
\end{equation*} 
The two contributions correspond respectively to the intra-layer and the inter-layer supra-Laplacians. The first one is a block-diagonal matrix where the generic block $\alpha$ is the standard Laplacian matrix of the individual layer $\alpha$, defined as:  $L^{[\alpha]}_{ij} = k^{[\alpha]}_i\delta_{ij} - a^{[\alpha]}_{ij}$, where $k_i^{[\alpha]} = \sum_{j} a^{[\alpha]}_{ij}$ denotes the degree of node $i$ at layer $\alpha$, and $\delta_{ij}$ is the Kronecker delta. Hence, we have:
\begin{equation*}
\Lp^{\ell}=
\left(
\begin{array}{c c c c}
 D^{[1]} L^{[1]} & & & \\ 
 & D^{[2]}  L^{[2]} & & \\
 & & \ddots& \\
 & & & D^{[M]} L^{[M]}
\end{array}
\right)
\end{equation*}

The inter-layer supra-Laplacian is instead a matrix composed by $M\times M$ blocks, each one being a diagonal matrix of dimension $N\times N$:

\begin{equation*}
\Lp^x=
\left(
\begin{array}{c c c c}
\sum_{\alpha}D_x^{[1,\alpha]}\bd{I}_N & -D_x^{[1,2]}\bd{I}_N &\hdots& -D_x^{[1,M]}\bd{I}_N\\ 
-D_x^{[2,1]}\bd{I}_N &\sum_{\alpha}D_x^{[2,\alpha]}\bd{I}_N & \hdots& -D_x^{[2,M]}\bd{I}_N\\ 
\vdots & &\ddots & \\
-D_x^{[M,1]}\bd{I}_N & & &\sum_{\alpha}D_x^{[M,\alpha]}\bd{I}_N 
\end{array}
\right),
\end{equation*}
where $\bd{I}_N$ is the identity matrix of dimension $N$. For simplicity, we will only consider processes where the intensity of diffusion between two layers is equal in both directions, i.e. $D_x^{[\alpha,\beta]}=D_x^{[\beta,\alpha]}$.

The solution of the system of equations~\eqref{diff} is found by exploiting the eigenvector basis $\phi^{(k)}$ of the supra-Laplacian $\Lp^{\mathcal M}$, which allows to express in exponential form the time evolution of the state of the system as: $\bd x(t) = \sum_{k} \exp(-\lambda_k^{\mathcal M}t) \bd \phi^{(k)}$, with $\{\lambda_k\}$ the set of supra-Laplacian eigenvalues. The analysis of the process is thus reduced to studying the supra-Laplacian spectrum, and the convergence to the equilibrium represented by the homogeneous state is governed by the smallest non-zero eigenvalue, $\lambda^{\mathcal M}_2$, also called {\it algebraic connectivity}~\cite{Fiedler73}, which sets the time-scale for the process.

While $\lambda^{\mathcal M}_2$ can in general be computed numerically, an approximate solution has been found in the two limits of strong and weak inter-layer coupling~\cite{Gomez_etal13}. Let us consider $D^{[\alpha]}$ of order one $\forall \alpha$, we have that:

(\textit{i}) in the limit $D_x^{[\alpha,\beta]}\ll1$, the smallest eigenvalues coincide with those of the inter-layer supra-Laplacian, so the algebraic connectivity is given by $\lambda^{\mathcal M}_2 = \lambda_2({\Lp}^x)$. In this limit, indeed, the bottleneck for the diffusion process is given by the weak connections between layers. This ultimately implies that the time scale of the diffusion process is set by inter-layer diffusion and does not depend on the structure of the individual layers.

(\textit{ii}) For $D_x^{[\alpha,\beta]}\gg1$, the smallest $N-1$ non-zero eigenvalues of $\Lp^{\mathcal M}$ can be approximated as the non-zero eigenvalues of the average single-layer Laplacian $L^{AV} = \frac{1}{M}\sum_{\alpha}L^{[\alpha]}$, i.e. $\lambda^{\mathcal M}_2 \approx  \lambda_2(L^{AV})$. Here the intra-layer networks come into play bringing with them all their topological features, which can promote or hinder the diffusion process. 

In general, the diffusion time scale of the multiplex network $\tau^{\mathcal M} \sim 1 / \lambda^{\mathcal M}_2$ is different from those of its individual layers in isolation $\tau^{[\alpha]} \sim 1 / \lambda^{[\alpha]}_2$, and the intensity of diffusion across layers can dramatically affect the velocity of the process. Interestingly, it has been found that multiplexity can lead to \textit{super-diffusion}, i.e. when the multiplex time scale is smaller than that of each layer, meaning that inter-layer diffusion facilitate the diffusion process~\cite{Gomez_etal13}.

We therefore define a super-diffusion indicator, which will be useful for our investigations, as the relative value of the first non-zero eigenvalue of the supra-Laplacian of the multiplex compared to that of the fastest layer:
\begin{equation}
\zeta = \frac{\lambda^{\mathcal M}_2 - \max_{\alpha}( \lambda^{[\alpha]}_2)}{ \max_{\alpha}( \lambda^{[\alpha]}_2)}.
\label{eq:zeta}
\end{equation}
If the multiplex outperforms the single layers we have super-diffusion and $\zeta>0$. We remark that, given a networked system, such indicator does not provide a  description of multiplex diffusion in absolute terms, but relatively to the performance of the individual layers. For this reason, having $\zeta_1>\zeta_2$ for two structurally distinct systems does not necessarily imply that diffusion is faster in the former, but rather that multiplexity has a more beneficial effect in this network compared to the latter. In the following we provide a wide overview of some key structural and dynamical determinants of the diffusive behavior of multiplex networks, unveiling and clarifying the emergence and the intensity of super-diffusion for a wide class of systems.


\begin{figure*}[ht]
\centering
\includegraphics[width=1.0\textwidth]{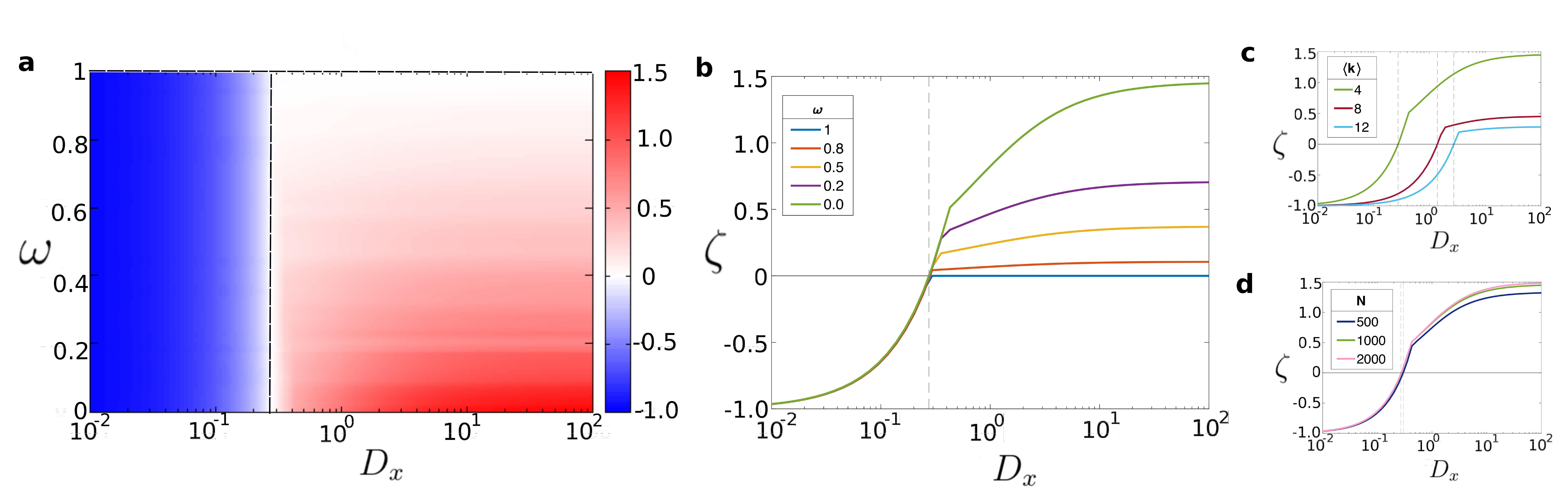}
\caption{\label{fig:fig1} \textbf{Structural determinants of multiplex diffusion in regular random networks.}  
(\textbf{a}) Super-diffusion index $\zeta$ as a function of the inter-layer diffusion coefficient $D_x$ (in logarithmic scale) and of the overlap $\omega$ for multiplex networks composed by two regular random graphs with average degree $k=4$ and $N=1000$ nodes, where the intra-layer diffusion coefficients $D^{[1]}$ and $D^{[2]} $ have been set equal to one. The onset of super-diffusion is independent from $\omega$, while its intensity is maximized in absence of overlap. The critical value for super-diffusion is $D_{x,c} =0.28$. (\textbf{b}) $\zeta$ as a function of $D_x$ for five selected values of overlap. In the non super-diffusive regime, $\zeta$ is independent from $\omega$. Multiplex super-diffusion is harder for increasing values of degree $k$ (\textbf{c}), while it is not strongly dependent on the system size $N$ (\textbf{d}).}
\end{figure*}

\section*{Results}
\label{sec_results}

{\bf Structural determinants of multiplex diffusion.}

We start our investigation by considering multiplex networks composed by two layers, with $D_x^{[1,2]}=D_x^{[2,1]}=D_x$. The structural correlation between two layers can be quantified by the edge overlap $\omega$~\cite{Battiston14, Bianconi13}:
\begin{equation}
\omega=\frac{\sum_{i,j>i}a_{ij}^{[1]}a_{ij}^{[2]}}{\sum_{i,j>i}(a_{ij}^{[1]} + a_{ij}^{[2]} - a_{ij}^{[1]}a_{ij}^{[2]})},
\label{eq:over2layers}
\end{equation}
measuring the fraction of connected pairs $i$ and $j$ which are linked at both layers, $0 \le \omega \le 1$. When $\omega=1$ the two networks are equivalent, hence in the limit $D_x\gg1$ we have that $\lambda^{\mathcal M}_2 =  \lambda_2^{[1]} = \lambda_2^{[2]}$ and $\zeta=0$. As in undirected multiplex networks $\lambda^{\mathcal M}_2$ is an increasing function of $D_x$, for weaker inter-layer diffusion $\zeta<0$ and super-diffusion can never be achieved. 
The intensity of super-diffusive behavior has been linked to high dissimilarity across layers in the past~\cite{Serrano_etal17}. Here we systematically investigate the effect of the overlap in controlled settings where we can tune at will the structural correlations across the layers of the multiplex, focusing both on the maximum achievable super-diffusion, as well as its onset.

In Figure~\ref{fig:fig1}a we show the super-diffusion index $\zeta$ as a function of both the overlap $\omega$ and the inter-layer coefficient $D_x$ for multiplex networks with $N=1000$ nodes composed by two regular random graphs (RRG) with the same degree $k^{[1]}=k^{[2]}=k=4$. Given two identical layers, it is possible to tune at will the value of edge overlap by rewiring a fraction of edges $f$ in one of the layers so that $\omega=(1-f)/(1+f)$~\cite{Diakonova16, BattistonPercLatora17}. The lack of structure in RRGs layers is functional to quantify the role of multiplex correlations for super-diffusion in a set up where intra-layer effects are minimized. Intra-layer diffusion coefficients $D^{[1]}$ and $D^{[2]}$ are set equal to $1$ without loss of generality. This figure and the following ones have been obtained by averaging over 50 independent realisation of a multiplex network with the indicated features.

As shown, as long as $\omega < 1$, multiplex super-diffusion occurs in the system, and, surprisingly, in this simple scenario its onset does not depend on the overlap. The super-diffusive regime (in red) is indeed only triggered by values of the inter-layer diffusion coefficient higher than a critical value $D_{x,c} \approx 0.28$. The onset of super-diffusion is independent from $\omega$ also for different choices of size $N$ and connectivity $k$ (results not shown). However, the overlap is still important to determine the intensity of super-diffusion: in the region $D_x>D_{x,c}$ we observe a clear gradient indicating that high values of $\zeta$ can only be obtained for small overlap, with a maximum of $\zeta_{\rm{max}}=1.45$ when $\omega=0$. Super-diffusion is hence maximized by minimizing the similarity in the structure of the layers, in this way also reducing the overall robustness of the system, as no pair of nodes is connected across multiple layers~\cite{Cellai13}.

In contrast, when the system is not super-diffusive, i.e. $\zeta<0$ (in blue), multiplex diffusion is independent from $\omega$. This can be easily observed in Figure~\ref{fig:fig1}b, where we report $\zeta$ as a function of $D_x$ for a few selected values of overlap. This is in agreement with the previously mentioned limit for $D_x^{[\alpha,\beta]}\ll1$, according to which the algebraic connectivity of the multiplex does not explicitly depend on the structure of layers, nor their overlap.
The lack of effect of $\omega$ for $D_x<D_{x,c}$ is also reminiscent of the existence of structural transitions in multiplex networks~\cite{RadicchiArenas13}, meaning that sufficiently strong values of the inter-layer links are needed for the system to feel as a whole and give rise to multiplex emergent dynamics.

\begin{figure*}[ht]
\centering
\includegraphics[width=1.0\textwidth]{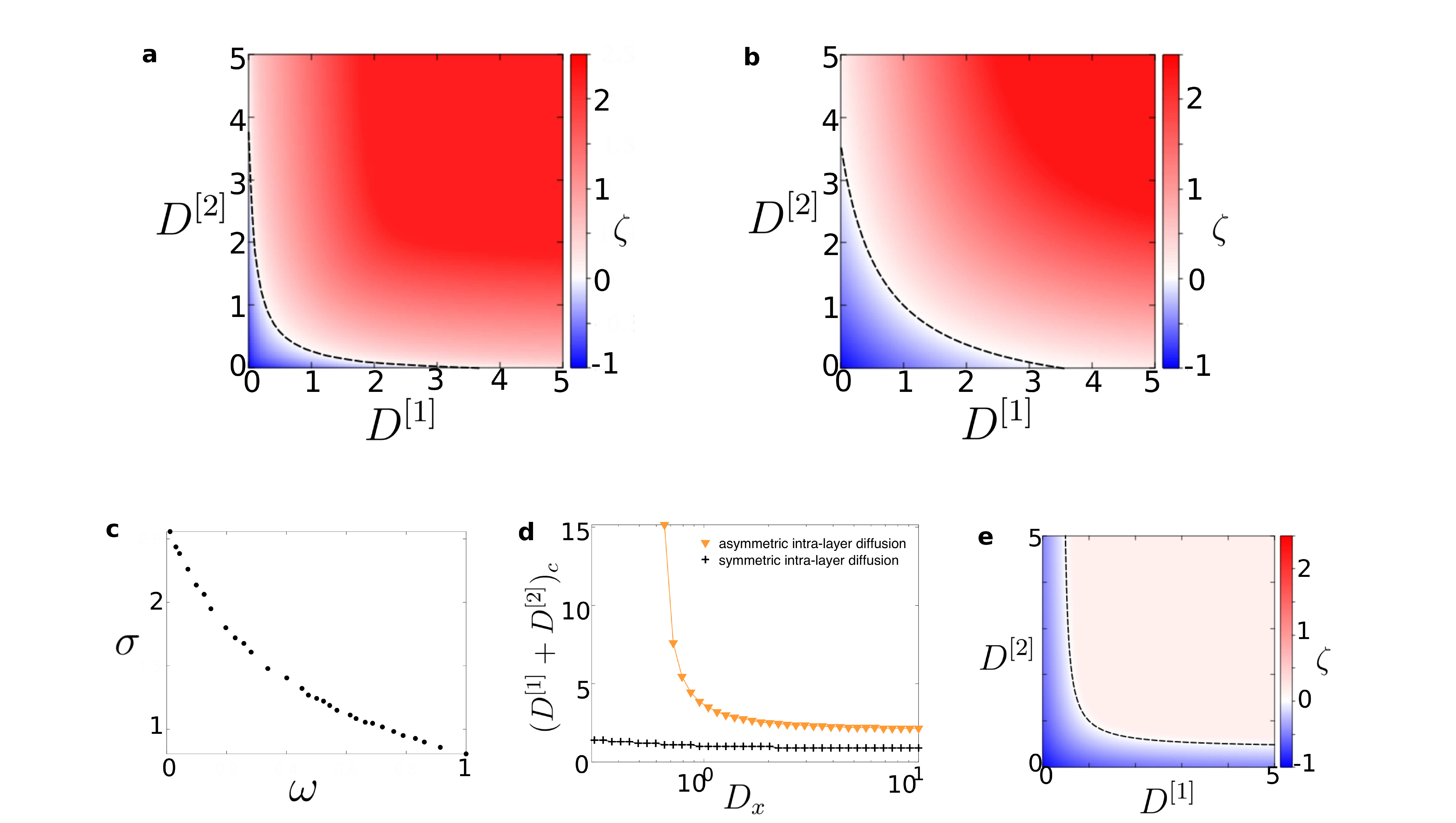}
\caption{\label{fig:fig2} \textbf{Dynamical determinants of multiplex diffusion in regular random networks.} 
Phase diagrams of $\zeta$ as a function of the two intra-layer diffusion coefficients $D^{[1]}$ and $D^{[2]}$ for $\omega=0$ (\textbf{a}) and $\omega=1$ (\textbf{b})
for regular random graphs with $k=4$ and $D_x=1$. Symmetric configurations with $D^{[1]} = D^{[2]}$ are the most convenient for the onset of super-diffusion, though the difference is the greatest for $\omega=0$. This effect can be quantified by $\sigma$, accounting for the relative additional cost associated to achieve super-diffusion in asymmetric dynamical configurations where $D^{[1]} \neq D^{[2]}$, shown as a function of $\omega$ (\textbf{c}). While in symmetric configurations super-diffusion is always achievable, this is not true for asymmetric configurations for values of inter-layer diffusion $D_x$ lower than 0.60 (\textbf{d}). As an example, we report $\zeta$ as a function of $D^{[1]}$ and $D^{[2]}$ for $\omega=1$ and $D_x=0.28$ (\textbf{e}).}
\end{figure*}

In Figures~\ref{fig:fig1}c,d,  we investigate also the effect of additional structural features of the system, namely the degree $k$ of the RRGs and the number of nodes $N$ in the system. As shown, the lower the $k$ the sooner the onset of super-diffusion, and the higher its intensity. At a first glance this result might seem counter-intuitive, as in single-layer networks the algebraic connectivity is typically an increasing function of the mean degree~\cite{VanMieghem10}. However, this means that, when coupling denser layers with higher intra-layer diffusions, larger inter-layer values of $D_x$ are necessary for multiplexity to be beneficial to diffusion in the system. 
In contrast, the size of the system does not affect significantly multiplex diffusion. This suggests that in real-world multiplex transportation networks, the addition of new nodes into the system might harm the beneficial effect of multiplexity, driving the networks outside the super-diffusive regime, unless the number of new links is adequately well-connected to the existing infrastructure. Results in these last two figures are shown for $\omega=0$, but similar effects are observed for all values of $\omega<1$.\\

{\bf Dynamical determinants of multiplex diffusion.}

Having clarified the basic structural determinants of multiplex diffusion, we now focus on the joint effect of the three dynamical parameters of the model, namely the inter-layer diffusion coefficients $D_x$, governing the intensity of the flow across layers, and the two intra-layer diffusion coefficients $D^{[1]}$ and $D^{[2]}$ governing flows within layers. 

In Figure~\ref{fig:fig2}a we show the super-diffusion index $\zeta$ as a function of $D^{[1]}$ and $D^{[2]}$ for a multiplex network composed by two RRGs with $k=4$ and $\omega=0$, and where we set $D_x=1$. The diffusion process is clearly facilitated by high values of $D^{[1]}$ and $D^{[2]}$, as expected. It is interesting to notice that multiplex networks corresponding to the same value of the sum $D^{[1]} + D^{[2]}$, i.e. characterized by on average the same intra-layer diffusion across the two layers, are not associated to the same diffusion time scale. In particular, $\zeta$ decreases while the difference between $D^{[1]}$ and $D^{[2]}$ increases and multiplex diffusion is maximized when the two coefficients are identical. In the case of real-world infrastructures, where we can assume that intra-layer coefficients $D^{[\alpha]}$ reflect the size of channels connecting nodes at the different layers (whose building cost is the same), this means that resources should not be allocated in a way to preferentially facilitate diffusion in one of the layers, as this could not simply slow down the system, but even carry the multiplex out of the super-diffusive phase.

In Figure~\ref{fig:fig2}b we show $\zeta$ for a multiplex network with maximum overlap $\omega=1$. As shown, because of the correlated structure of the two layers, super-diffusion generally emerges at higher values of intra-layer coefficients. Also in the limit where one of the two coefficients, say $D^{[1]}$, tends to zero, interestingly super-diffusion is still achievable as long as $D^{[2]}$ is greater than a critical value ($D^{[2]}_c \approx 3.6$ for the system under consideration). Interestingly such value is independent from $\omega$.
We refer to this latter case as the maximally asymmetric dynamic configuration $\mathcal A_{\rm{max}}$ that sustains super-diffusion. Conversely, we indicate the case $D^{[1]} = D^{[2]} = D$ as a dynamically symmetric configuration $\mathcal S$. We quantify the relative additional cost to achieve super-diffusion in asymmetric configurations by the index $\sigma$, which accounts for the relative difference between the critical value of the sum of the intra-layer diffusion coefficients for the maximally asymmetric and symmetric configurations. In formulae
\begin{equation}
\sigma = \frac{(D^{[1]} + D^{[2]})_{c,{\mathcal{A}_{\rm{max}}}} - (D^{[1]} + D^{[2]})_{c,{\mathcal{S}}}}{(D^{[1]} + D^{[2]})_{c,{\mathcal{S}}}},
\end{equation}
where $\sigma=0$ indicates that there is no additional cost for multiplex super-diffusion in asymmetric dynamical configurations. As shown in Figure~\ref{fig:fig2}c, the higher the overlap the lower the relative cost $\sigma$, with a minimum of $\sigma_{\rm{min}} \approx 0.78$ for $\omega=1$. In contrast, for $\omega=0$ the increased cost of the asymmetric configurations are maximum, and $\sigma_{\rm{max}} \approx 2.65$.  

Interestingly, in this case where $D_x=1$, multiplex super-diffusion is always achievable also in such asymmetric configurations. However, this is not always true.
In Figure~\ref{fig:fig2}d we show the critical value of intra-layer diffusions for both $\mathcal S$ (black crosses) and $\mathcal A_{\rm{max}}$ (orange triangles) as a function of $D_x$. In general, the lower $D_x$ the harder it is to achieve super-diffusion. More importantly, for values of $D_x$ smaller than $D_{x_c}\approx 0.6$ $D^{[2]}_c (D^{[1]} \to 0)$ diverges, meaning that super-diffusion can not be achieved if contributions to multiplex diffusion from the two layers are very unbalanced, independently from the value of overlap. As an example, in Figure~\ref{fig:fig2}e we show $\zeta$ as a function of $D^{[1]}$ and $D^{[2]}$ for $\omega=1$ and $D_x=0.28$. In contrast, symmetric configurations always allow to reach super-diffusion, and for this reason are to be preferred.\\

\begin{figure*}[ht]
\centering
\includegraphics[width=1.0\textwidth]{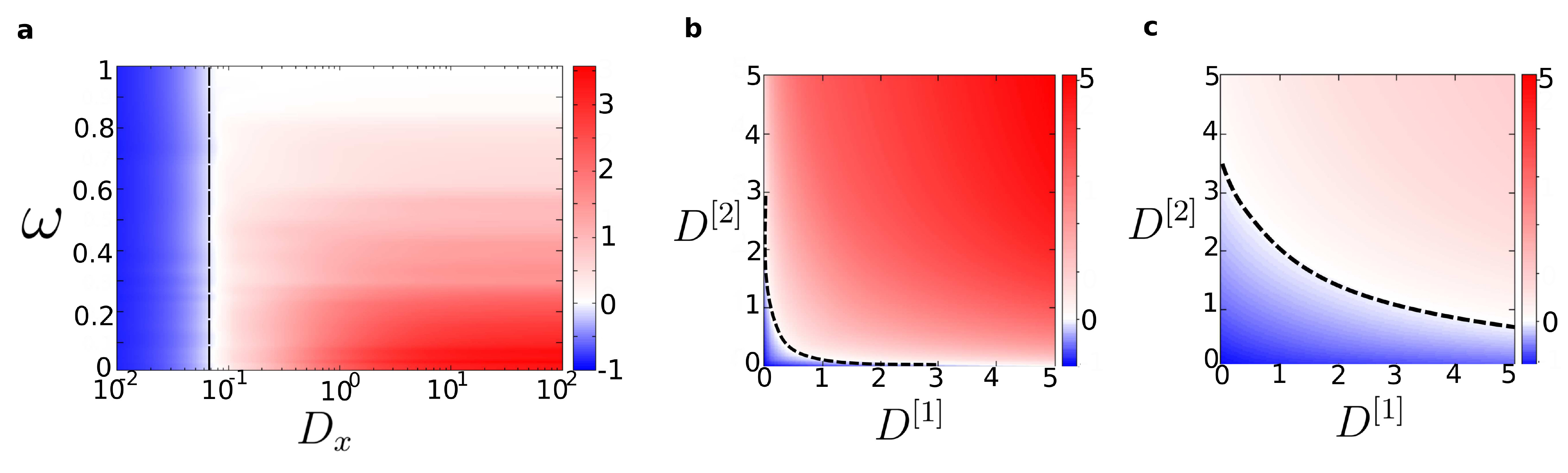}
\caption{\label{fig:fig3} \textbf{Multiplex super-diffusion in heterogenous networks.}  (\textbf{a}) Super-diffusion index $\zeta$ as a function of the inter-layer diffusion coefficient $D_x$ (in logarithmic scale) and of the overlap $\omega$ for multiplex networks composed by two SF networks with average degree $\langle k \rangle=4$, $N=1000$ and $\gamma = 2.5$. In the limit of large $D_x$ the intensity of super-diffusion is maximum for low overlap, whereas its onset is independent from $\omega$ and occurs at $D_{x,c} \approx 0.08$, a value smaller than the corresponding scenario for regular random graphs.} Phase diagrams of $\zeta$ as a function of $D^{[1]}$ and $D^{[2]}$ ($D_x=1$) for two layers with $\omega=0$ and (\textbf{b})  $\gamma^{[1]} = \gamma^{[2]} = 2.5$, (\textbf{c})  $\gamma^{[1]} = 2.8$, $\gamma^{[2]} = 2.2$. The first symmetric case facilitates super-diffusion. In the latter case, super-diffusion is more easily achieved by promoting diffusion in the layer with the broadest degree distribution.
\end{figure*}

{\bf Multiplex super-diffusion in heterogenous networks.}
For the sake of simplicity, in the first two sections we focused on the effects of structural correlations and diffusion coefficients on multiplex diffusion by considering multiplex networks composed by RRGs, where all nodes have the same degree $k$. However, the layers of many real-world systems are typically characterised by broad degree distributions. In this Section we investigate multiplex diffusion on scale-free (SF) networks, characterised by power-law degree distributions $p(k)\simeq k^{-\gamma}$. 

In Figure~\ref{fig:fig3}a we show the super-diffusion index $\zeta$ as a function of both $D_x$ and $\omega$ for two SF networks, both of them with $\gamma=2.5$ and $\langle k \rangle=4$, where we set $D^{[1]}=D^{[2]}=1$. Similarly to the case of homogenous layers, in the limit of large $D_x$ super-diffusion is maximized in absence of overlap. Besides, the onset of the super-diffusive regime is independent from $\omega$. We numerically estimated the critical value of inter-layer diffusion at $D_{x,c} \approx 0.08$, a value smaller than that obtained for regular random graphs. As for sparser layers, networks with a broad degree distribution typically have a smaller algebraic connectivity, associated to slower intra-layer diffusion, meaning that smaller inter-layer values of $D_x$ are sufficient for diffusion to benefit from the interconnected nature of the system. We remark that the rewiring process used to generate multiplex networks with different values of overlap preserves the original degree sequence. Thus, all considered configurations here have inter-layer degree correlations maximum and equal to 1~\cite{nicosia2015}. For a fixed overlap, say $\omega=0$, the absence of inter-layer degree correlations fosters the emergence of super-diffusion (results not shown).

In Figure~\ref{fig:fig3}b we show $\zeta$ as a function of $D^{[1]}$ and $D^{[2]}$ for $\omega=0$ and $D_x=1$. High values of intra-layer coefficients facilitate diffusion, which is maximized for $D^{[1]} = D^{[2]}$. Compared to the analogous case for RRGs shown in Figure~\ref{fig:fig2}a, dynamical asymmetry in intra-layer diffusion is less penalizing in SF networks, as we find $\sigma_{\rm{SF}}(\omega=0)=1.69 < \sigma_{\rm{RRG}}(\omega=0)=2.65$. Finally, in Figure~\ref{fig:fig3}c we show $\zeta$ for two SF layers with $\omega=0$, $\langle k\rangle=4$ and different power-law exponents, namely $\gamma^{[1]} = 2.8$ and $\gamma^{[2]} = 2.2$. First, we notice that the structural asymmetry does not help the the relative velocity of the diffusion process, nor the onset of multiplex super-diffusion. In both Figure~\ref{fig:fig3}b and Figure~\ref{fig:fig3}c, we considered multiplex networks without inter-layer degree correlations.

Differently from before, because of the structural asymmetry in the system, for constant values of $D^{[1]} + D^{[2]}$ super-diffusion is not maximized for $D^{[1]} = D^{[2]}$. As $\gamma^{[2]} < \gamma^{[1]}$, the second layer has a broader degree distribution and is diffusing faster than the first one. Consequently, if increasing the diffusive potential of the two layers has the same cost, it is preferable to allocate resources on the second layer.\\

\begin{figure*}[ht]
\centering
\includegraphics[width=1.0\textwidth]{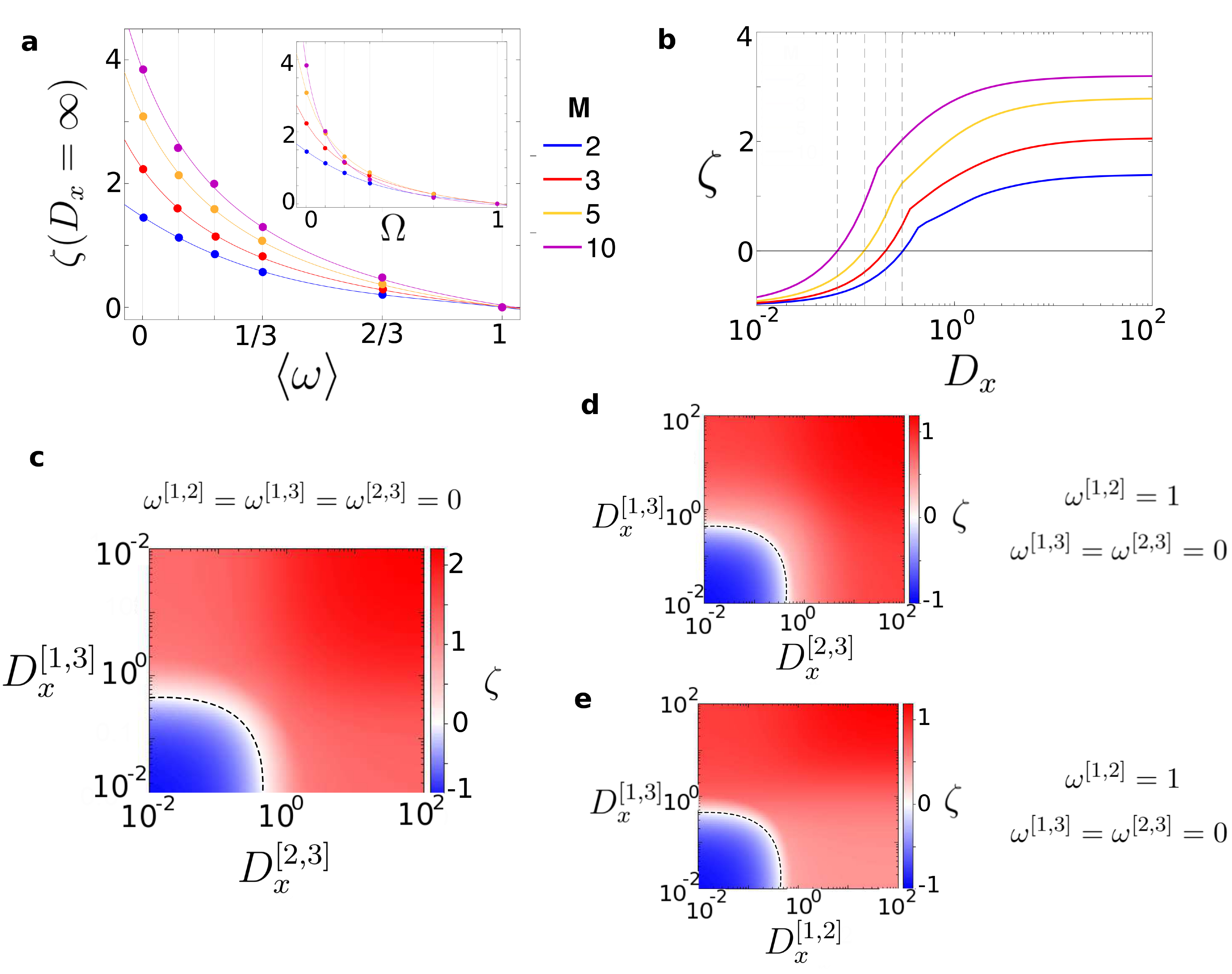}
\caption{\label{fig:fig4} \textbf{Multiplex super-diffusion in networks with many layers.} (\textbf{a}) Super-diffusion index $\zeta$ for multiplex networks made of different number of layers as a function of two generalized measures of overlap $\langle \omega \rangle$ and $\Omega$ (inset). The higher the number of layers, the stronger the super-diffusion, and the earlier its onset, as shown in (\textbf{b}) in absence of overlap. For $M=3$, phase diagram of $\zeta$ as a function of different inter-layer diffusion coefficients for completely uncorrelated layers, i.e. $\langle\omega\rangle=0$ (\textbf{c}). For $\omega^{[1,2]}=1$, $\omega^{[1,3]}=\omega^{[2,3]}=0$, $\zeta$ as a function of different inter-layer diffusion coefficients (\textbf{d} and \textbf{e}). In all three cases the onset of super-diffusion is not affected by the overlap. For partially overlapping layers, multiplex diffusion is best enhanced when promoting diffusion across different layers.
}
\end{figure*}

{\bf Multiplex super-diffusion in networks with many layers.}
All the above considerations can be easily extended to multiplex networks formed by many layers, i.e. $M>2$. 
In this type of system, the overlap is often quantified as the average overlap~\cite{BattistonPercLatora17}
\begin{equation}
\langle \omega \rangle = \frac{2}{M(M-1)} \sum_{\alpha,\beta>\alpha} \omega^{[\alpha,\beta]},
\end{equation}
where $\omega^{[\alpha,\beta]}$ is the overlap between layers $\alpha$ and $\beta$.
In Figure~\ref{fig:fig4}a we report the super-diffusion index $\zeta$ as a function of $\langle \omega \rangle$ in the limit of large $D_x$ for multiplex networks formed by different number of layers $M$, each one of them an RRG with $\langle k \rangle = 4$. As shown, for the same mean overlap, multiplex networks with more layers are able to achieve a higher multiplex super-diffusion. In the inset of Figure~\ref{fig:fig4}a, we report $\zeta$ as a function of a different measure of aggregate overlap, defined as
\begin{equation}
\Omega = \frac{\langle e \rangle - 1}{M - 1}
\end{equation}
where 
\begin{equation}
\langle e \rangle = \frac{  \sum_{i, j > i} \sum_{\alpha} a^{[\alpha]}_{ij} }{\sum_{i, j > i} 1 - \delta_{0,\sum_{\alpha} a^{[\alpha]}_{ij} }.  }
\end{equation}
is the mean number of layers at which connected pairs of nodes in the multiplex networks are linked. Similarly to the mean overlap, $0 \le \Omega \le 1$, and the measure reduces to Eq.~\ref{eq:over2layers} for a system with two layers only. Differently from the mean overlap, however, this measure of overlap is not limited to capturing structural correlations between two layers at a time. As shown, curves for different number of layers appear to collapse, at least for sufficiently high values of aggregate overlap $\Omega > 1/3$. This hints at the necessity to investigate the effect of higher-order non-pairwise layer correlations for dynamical processes on multiplex networks, a direction that we leave for future work. In Figure~\ref{fig:fig4}b, we report $\zeta$ for $\langle \omega \rangle = \Omega = 0$ as a function of $D_x$. As shown, a higher number of layer not only facilitates the intensity of super-diffusion, but also its onset.

Finally, we investigate the effect of heterogenous inter-layer diffusion coefficients $D_x^{[\alpha, \beta]}$ in multiplex networks. For the sake of simplicity we show results for $M=3$, all RRGs with $\avk=4$, but the following findings can be easily extended to the case of networks with a generic number of layers. In Figure~\ref{fig:fig4}c we show $\zeta$ as a function of $D_x^{[1,3]}$ and $D_x^{[2,3]}$ (in log-log scale), for a system with $\langle \omega \rangle = \Omega = 0$, $D^{[1]} =D^{[2]} = D^{[3]} =1$, and where we set $D_x^{[1,2]}=1$. Once again diffusion is maximally speeded up when the coefficients assume their highest values. As for the case of intra-layer diffusion, by plotting the same figure in linear scale it appears that symmetric configurations are slightly preferred under the constraint that the sum $D_x^{[1,3]}+D_x^{[2,3]}$ is constant (plot not shown).

In Figure~\ref{fig:fig4}d multiplex diffusion is investigated for a system where the first two layers are identical and completely different from the third one, i.e. $\omega^{[1,2]} = 1$,  $\omega^{[1,3]} = \omega^{[2,3]} = 0$. By changing the values of $D_x^{[1,3]}$ and $D_x^{[2,3]}$ (again we set $D_x^{[1,2]}=1$), in both cases we are tuning inter-layer diffusion between non-overlapping layers, hence as expected the phase diagram is symmetrical, describing inter-layer diffusion between identical layers, is again set to 1). Remarkably, the region of non-superdiffusivity in the phase diagram for Figure~\ref{fig:fig4}c and Figure~\ref{fig:fig4}d is the same, even though the two multiplex networks have very different overlap. This can be considered as a generalization of the findings of Figures~\ref{fig:fig1}a,b to the case of networks with arbitrary number of homogenous layers, for which the onset of super-diffusion is independent from $\langle \omega \rangle$. In contrast, the overlap affects the intensity of super-diffusion, which is maximized when the structure of the layers is maximally different (the colorbars of Figure~\ref{fig:fig4}c and Figure~\ref{fig:fig4}d have different scales). Last, in Figure~\ref{fig:fig4}e we show $\zeta$ for the same multiplex network, this time as function of $D_x^{[1,2]}$ (inter-layer diffusion across identical layers) and $D_x^{[1,3]}$ (inter-layer diffusion across different layers. We set the other coefficient $D_x^{[2,3]}=1$). As the phase diagram shows, $D_x^{[1,3]}$ is responsible for a much greater variability of $\zeta$ than the other coefficient. This means that multiplex diffusion is promoted more by enhancing diffusion across different layers, rather than similar ones.\\

\section*{Conclusions}	
Diffusion was one of the first dynamics to be investigated on networks with multiple layers of interactions. In particular multiplex super-diffusion is a paradigmatic example of novel emergent behavior which can not be understood by considering each layer in isolation or by merging together the different networks of interactions. In Ref.~\cite{Gomez_etal13} selected network topologies were shown to sustain super-diffusion, even though no general theory for its emergence was derived. 
In this work we have clarified the structural and dynamical determinants of super-diffusive behavior in multiplex networks. We have shown that, for large diffusion within layers, the absence of overlap is crucial to maximize the beneficial effect of the multiplex structure to the process. However, low overlap is not favourable to the system for low and intermediate values of inter-layer diffusion.
In fact, surprisingly, the onset of super-diffusion is independent from the presence of structural correlations, both for regular random and scale-free layers. We have also shown that, when building costly channels within each layer, an equal allocation of resources is preferable, as balanced intra-layer diffusion best sustains multiplex diffusion. More importantly, an unequal allocation across layers might cause the system to jump out of the super-diffusion regime, thus eliminating the potential beneficial effect of multiplexity. Last, for systems composed by a large number of layers, our analysis suggests that pairwise network distance is not enough to fully capture the complexity of the process. For such reason, we urge that more efforts be focused to link the emergence of multiplex dynamics to the underlying structure measured at a global scale, as by multiplex reducibility~\cite{DeDomenico15}. 

In conclusion, our work sheds new light on the diffusive behavior of multiplex networks, and hopefully will trigger further investigations in this directions. In particular, combining real-world multiplex properties with realistic directed diffusion within layers, recently associated to the emergence of a new super-diffusive regime~\cite{Tejedor_etal18}, seems to us a particularly promising direction to explore, that we leave for the future.

\section*{Acknowledgements}
The authors thank Iacopo Iacopini for carefully reading and providing suggestions on the manuscript.\\
GC acknowledges COSTNET for the financial support and the Department of Network and Data Sciences at Central European University for the hospitality.

	\addcontentsline{toc}{chapter}{Bibliography}
	\bibliographystyle{ieeetr} 
	\bibliography{bib_multiplex}

\begin{thebibliography}{10}

\bibitem{Gardiner09}
C.~Gardiner, {\em Stochastic methods}, vol.~4.
\newblock springer Berlin, 2009.

\bibitem{Crank79}
J.~Crank {\em et~al.}, {\em The mathematics of diffusion}.
\newblock Oxford university press, 1979.

\bibitem{Murray02}
J.~D. Murray, ``Mathematical biology i: an introduction, vol. 17 of
  interdisciplinary applied mathematics,'' 2002.

\bibitem{Turing52}
A.~M. Turing, ``The chemical basis of morphogenesis,'' {\em Philosophical
  Transactions of the Royal Society of London B: Biological Sciences},
  vol.~237, no.~641, pp.~37--72, 1952.

\bibitem{Boccaletti06}
S.~Boccaletti, V.~Latora, Y.~Moreno, M.~Chavez, and D.-U. Hwang, ``Complex
  networks: Structure and dynamics,'' {\em Physics reports}, vol.~424, no.~4-5,
  pp.~175--308, 2006.

\bibitem{BarratBarthelemyVespignani08}
A.~Barrat, M.~Barthelemy, and A.~Vespignani, {\em Dynamical processes on
  complex networks}.
\newblock Cambridge university press, 2008.

\bibitem{MasudaPorterLambiotte17}
N.~Masuda, M.~A. Porter, and R.~Lambiotte, ``Random walks and diffusion on
  networks,'' {\em Physics Reports}, 2017.

\bibitem{Strogatz01}
S.~H. Strogatz, ``Exploring complex networks,'' {\em nature}, vol.~410,
  no.~6825, p.~268, 2001.

\bibitem{reactRW}
G.~Cencetti, F.~Battiston, D.~Fanelli, and V.~Latora, ``Reactive random walkers
  on complex networks,'' {\em Phys. Rev. E}, vol.~98, p.~052302, Nov 2018.

\bibitem{Szell10}
M.~Szell, R.~Lambiotte, and S.~Thurner, ``Multirelational organization of
  large-scale social networks in an online world,'' {\em Proceedings of the
  National Academy of Sciences}, vol.~107, no.~31, pp.~13636--13641, 2010.

\bibitem{DeDomenico14navigability}
M.~De~Domenico, A.~Sol{\'e}-Ribalta, S.~G{\'o}mez, and A.~Arenas,
  ``Navigability of interconnected networks under random failures,'' {\em
  Proceedings of the National Academy of Sciences}, vol.~111, no.~23,
  pp.~8351--8356, 2014.

\bibitem{Boccaletti_etal14review}
S.~Boccaletti, G.~Bianconi, R.~Criado, C.~I. Del~Genio, J.~G{\'o}mez-Gardenes,
  M.~Romance, I.~Sendina-Nadal, Z.~Wang, and M.~Zanin, ``The structure and
  dynamics of multilayer networks,'' {\em Physics Reports}, vol.~544, no.~1,
  pp.~1--122, 2014.

\bibitem{Kivela_etal14review}
M.~Kivel{\"a}, A.~Arenas, M.~Barthelemy, J.~P. Gleeson, Y.~Moreno, and M.~A.
  Porter, ``Multilayer networks,'' {\em Journal of complex networks}, vol.~2,
  no.~3, pp.~203--271, 2014.

\bibitem{Battiston_etal17}
F.~Battiston, V.~Nicosia, and V.~Latora, ``The new challenges of multiplex
  networks: Measures and models,'' {\em The European Physical Journal Special
  Topics}, vol.~226, pp.~401--416, Feb 2017.

\bibitem{Gomez_etal13}
S.~Gomez, A.~Diaz-Guilera, J.~Gomez-Gardenes, C.~J. Perez-Vicente, Y.~Moreno,
  and A.~Arenas, ``Diffusion dynamics on multiplex networks,'' {\em Physical
  review letters}, vol.~110, no.~2, p.~028701, 2013.

\bibitem{Salehi_etal15}
M.~Salehi, R.~Sharma, M.~Marzolla, M.~Magnani, P.~Siyari, and D.~Montesi,
  ``Spreading processes in multilayer networks,'' {\em IEEE Transactions on
  Network Science and Engineering}, vol.~2, no.~2, pp.~65--83, 2015.

\bibitem{DeDomenico16}
M.~De~Domenico, C.~Granell, M.~A. Porter, and A.~Arenas, ``The physics of
  spreading processes in multilayer networks,'' {\em Nature Physics}, vol.~12,
  no.~10, p.~901, 2016.

\bibitem{deArruda18fundamentals}
G.~F. de~Arruda, F.~A. Rodrigues, and Y.~Moreno, ``Fundamentals of spreading
  processes in single and multilayer complex networks,'' {\em Physics Reports},
  2018.

\bibitem{Sole_etal13}
A.~Sole-Ribalta, M.~De~Domenico, N.~E. Kouvaris, A.~Diaz-Guilera, S.~Gomez, and
  A.~Arenas, ``Spectral properties of the laplacian of multiplex networks,''
  {\em Physical Review E}, vol.~88, no.~3, p.~032807, 2013.

\bibitem{RadicchiArenas13}
F.~Radicchi and A.~Arenas, ``Abrupt transition in the structural formation of
  interconnected networks,'' {\em Nature Physics}, vol.~9, no.~11, p.~717,
  2013.

\bibitem{Sanchez_etal14}
R.~J. S{\'a}nchez-Garc{\'\i}a, E.~Cozzo, and Y.~Moreno, ``Dimensionality
  reduction and spectral properties of multilayer networks,'' {\em Physical
  Review E}, vol.~89, no.~5, p.~052815, 2014.

\bibitem{Cozzo16spectral}
E.~Cozzo, G.~F. de~Arruda, F.~A. Rodrigues, and Y.~Moreno, ``Multilayer
  networks: metrics and spectral properties,'' in {\em Interconnected
  Networks}, pp.~17--35, Springer, 2016.

\bibitem{deArruda_etal18}
G.~F. de~Arruda, E.~Cozzo, F.~A. Rodrigues, and Y.~Moreno, ``A polynomial
  eigenvalue approach for multiplex networks,'' {\em New Journal of Physics},
  vol.~20, no.~9, p.~095004, 2018.

\bibitem{Asllani_etal14}
M.~Asllani, D.~M. Busiello, T.~Carletti, D.~Fanelli, and G.~Planchon, ``Turing
  patterns in multiplex networks,'' {\em Physical Review E}, vol.~90, no.~4,
  p.~042814, 2014.

\bibitem{KouvarisHataDiaz15}
N.~E. Kouvaris, S.~Hata, and A.~D{\'\i}az-Guilera, ``Pattern formation in
  multiplex networks,'' {\em Scientific reports}, vol.~5, p.~10840, 2015.

\bibitem{BusielloCarlettiFanelli18}
D.~M. Busiello, T.~Carletti, and D.~Fanelli, ``Homogeneous-per-layer patterns
  in multiplex networks,'' {\em EPL (Europhysics Letters)}, vol.~121, no.~4,
  p.~48006, 2018.

\bibitem{delGenio_etal16}
C.~I. del Genio, J.~G{\'o}mez-Garde{\~n}es, I.~Bonamassa, and S.~Boccaletti,
  ``Synchronization in networks with multiple interaction layers,'' {\em
  Science advances}, vol.~2, no.~11, p.~e1601679, 2016.

\bibitem{Serrano_etal17}
A.~B. Serrano, J.~G{\'o}mez-Garde{\~n}es, and R.~F. Andrade, ``Optimizing
  diffusion in multiplexes by maximizing layer dissimilarity,'' {\em Physical
  Review E}, vol.~95, no.~5, p.~052312, 2017.

\bibitem{Cellai13}
D.~Cellai, E.~L{\'o}pez, J.~Zhou, J.~P. Gleeson, and G.~Bianconi, ``Percolation
  in multiplex networks with overlap,'' {\em Physical Review E}, vol.~88,
  no.~5, p.~052811, 2013.

\bibitem{Battiston16efficient}
F.~Battiston, V.~Nicosia, and V.~Latora, ``Efficient exploration of multiplex
  networks,'' {\em New Journal of Physics}, vol.~18, no.~4, p.~043035, 2016.

\bibitem{Battiston17axelrod}
F.~Battiston, V.~Nicosia, V.~Latora, and M.~San~Miguel, ``Layered social
  influence promotes multiculturality in the axelrod model,'' {\em Scientific
  Reports}, vol.~7, no.~1, p.~1809, 2017.

\bibitem{BattistonPercLatora17}
F.~Battiston, M.~Perc, and V.~Latora, ``Determinants of public cooperation in
  multiplex networks,'' {\em New Journal of Physics}, vol.~19, no.~7,
  p.~073017, 2017.

\bibitem{Fiedler73}
M.~Fiedler, ``Algebraic connectivity of graphs,'' {\em Czechoslovak
  mathematical journal}, vol.~23, no.~2, pp.~298--305, 1973.

\bibitem{Battiston14}
F.~Battiston, V.~Nicosia, and V.~Latora, ``Structural measures for multiplex
  networks,'' {\em Phys. Rev. E}, vol.~89, p.~032804, Mar 2014.

\bibitem{Bianconi13}
G.~Bianconi, ``Statistical mechanics of multiplex networks: Entropy and
  overlap,'' {\em Physical Review E}, vol.~87, no.~6, p.~062806, 2013.

\bibitem{Diakonova16}
M.~Diakonova, V.~Nicosia, V.~Latora, and M.~San~Miguel, ``Irreducibility of
  multilayer network dynamics: the case of the voter model,'' {\em New Journal
  of Physics}, vol.~18, no.~2, p.~023010, 2016.

\bibitem{VanMieghem10}
P.~Van~Mieghem, {\em Graph spectra for complex networks}.
\newblock Cambridge University Press, 2010.

\bibitem{nicosia2015}
V.~Nicosia and V.~Latora, ``Measuring and modeling correlations in multiplex
  networks,'' {\em Physical Review E}, vol.~92, no.~3, p.~032805, 2015.

\bibitem{DeDomenico15}
M.~De~Domenico, V.~Nicosia, A.~Arenas, and V.~Latora, ``Structural reducibility
  of multilayer networks,'' {\em Nature communications}, vol.~6, p.~6864, 2015.

\bibitem{Tejedor_etal18}
A.~Tejedor, A.~Longjas, E.~Foufoula-Georgiou, T.~T. Georgiou, and Y.~Moreno,
  ``Diffusion dynamics and optimal coupling in multiplex networks with directed
  layers,'' {\em Physical Review X}, vol.~8, no.~3, p.~031071, 2018.

\end{thebibliography}

\end{document}